\documentclass[11pt,a4]{article}
\usepackage{amsmath}
\usepackage{epsfig}
\usepackage{amssymb}
\def\ra{ \rightarrow }
\def\cc{c\bar{c}}
\def\bb{b\bar{b}}
\def\be{\begin{equation}}
\def\ee{\end{equation}}

\newcommand{\epjc}[3]{{\it Eur.~Phys.~J.~}{\bf C #1} (#2) #3}

\newcommand{\jetp}[3]{{\it Sov.~Phys.~JETP }{\bf #1} (#2) #3}

\newcommand{\npb}[3]{{\it Nucl.~Phys.~}{\bf B #1} (#2) #3}

\newcommand{\plb}[3]{{\it Phys.~Lett.~}{\bf B #1} (#2) #3}
\newcommand{\prd}[3]{{\it Phys.~Rev.~}{\bf D #1} (#2) #3}
\newcommand{\prl}[3]{{\it Phys.~Rev.~Lett.~}{\bf #1} (#2) #3}

\newcommand{\sjnp}[3]{{\it Sov.~J.~Nucl.~Phys.~}{\bf #1} (#2) #3}
\newcommand{\zpc}[3]{{\it Z.~Phys.~}{\bf C #1} (#2) #3}
\begin{document}
\titlepage
\begin{flushright}    
DESY 03--171 \\    
October 2003    
\end{flushright}

\vspace*{1in}    
\begin{center}    
{\Large \bf Physics of ultrahigh energy neutrinos}\\    
\vspace*{0.4in}    
A.\ M. \ Sta\'sto \\    
\vspace*{0.5cm}  
{\it DESY, Theory Division, Notkestrasse 85, 22603 Hamburg, Germany} \\      
and \\
{\it Institute of Nuclear Physics, Radzikowskiego 152,    
 Krak\'ow, Poland}\\   

 \vskip 2mm    
\end{center}    
\vspace*{1cm}

\vskip1cm    
\begin{abstract}    
Ultrahigh energy neutrinos can provide important information about 
the distant astronomical objects and the origin of the Universe.
Precise knowledge about their interactions and production rates
is essential for estimating background, expected fluxes and detection probabilities.
 In this paper we review the
applications of the  high energy QCD
to the calculations of the interaction cross sections of the neutrinos.
We also study the production of the ultrahigh energy neutrinos in the atmosphere due to the charm and beauty decays.

\end{abstract}

\newpage
\section{Introduction}
The ultrahigh energy neutrino physics has attracted a lot of attention
during last couple of years. Neutrinos are important messengers of information
on distant extragalactic sources  in the Universe and  the neutrino astronomy
has many advantages over the conventional photon astronomy.
First of all, neutrinos, unlike photons, interact only weakly, so they can
travel long distances  being practically undisturbed. 
The typical interaction lengths for the neutrino and the photon at energy $E \sim 1 \, {\rm TeV}$ are about 
$$
{\cal L}_{int}^{\nu} \sim 250 \times 10^9 \, {\rm g/cm^2} \; , \hspace*{2cm}
{\cal L}_{int}^{\gamma} \sim 100 \, {\rm g/cm^2} \; .
$$
Thus very energetic photons with energy bigger than $ 10 \; {\rm TeV}$ cannot reach the Earth from the very distant corners
of our Universe without being rescattered. On the other hand neutrinos can travel very long distances and are also not deflected by  galactic magnetic fields.

At ultrahigh energies the angular distortion of the neutrino is very small.
Typically the angle between the neutrino and the produced muon is about 
$$
\delta \phi \simeq \frac{0.7^o}{(E_{\nu}/{\rm TeV})^{0.7}} \; .
$$
Therefore the highly energetic neutrinos  point back to their sources.
The interest in the  neutrinos at these high energies lead to the development
of the experimental devices such as AMANDA \cite{AMANDA},  NESTOR \cite{NESTOR}, ANTARES \cite{ANTARES} or planned ICE CUBE \cite{ICECUBE}
observatories.
For the reliable observation of the neutrinos, their  cross sections and  production rates in hadronic matter have to be well known. 
Even though the neutrinos interact only weakly with other particles, strong interactions play an essential role in the calculations of the neutrino production
rates and their interaction cross section. This is due to the fact that neutrinos are coming from the decays of various mesons such as $\pi, K, D$ and
even $B$ which are produced in the high energy proton-proton (or proton-nucleus, nucleus-nucleus) collisions. These hadronic processes occur mainly in the atmosphere
though possibly can be also present in the accretion disc in the remote
Active Galactic Nuclei. Also, the interactions of highly energetic neutrinos with matter are dominated by the deep inelastic cross section with the nucleons/nuclei as described in Sec.~4.
This is why the knowledge about QCD from high energy collider experiments such as HERA and TEVATRON is
invaluable.
The production and interaction of highly
energetic neutrinos with nucleons  involves gluon distribution probed at very small
values of Bjorken variable $x \, ( \sim 10^{-9})$ which corresponds to a very high c.m.s. energy. This is the kinematic domain which is currently inaccessible in these colliders (at HERA the minimal $x_{\rm min} \sim 10^{-5}$ at $Q^2 \sim 1 {\rm GeV^2}$, where $-Q^2$ is the virtuality of the photonic probe),
 and this calls for a reliable extrapolation of  parton densities
constrained by the current experimental data.
In this review we concentrate mainly on the
application of the knowledge and experience gained in testing high energy QCD  at present colliders to the
 problem of precise evaluation
of the neutrino-nucleon (or nucleus) scattering cross sections.
In addition, the production rate for the atmospheric neutrino flux from the
charm or beauty decays (the so called prompt neutrino flux) will be
analyzed.
 
This paper is organised as follows: in the next section we briefly discuss 
the origin of the ultrahigh energetic neutrinos. In Sec.~3 we describe the mechanism of the prompt neutrino production at high energies in the atmosphere and present the calculation of the cross section using  different models for the small $x$ dynamics, including
parton shadowing. We then follow the sequence of fragmentation, interaction and decays of the charmed mesons to obtain the neutrino fluxes at high energies. In Sec.~4 we discuss the interaction of the neutrinos with the nucleons and in particular we evaluate the deep inelastic neutrino-nucleon cross section based on the various parametrisations of the parton distribution functions.
In Sec.~5 we follow the process of the neutrino transport through the Earth
 and present the angular dependence of the fluxes after travelling
through the matter. Also in Sec.~5.1 we discuss the effect of the $\nu_{\tau}$
regeneration. Finally, in Sec.~6 we discuss the effects of the neutrino oscillations. 
\section{Sources of high energy neutrinos}

There are various sources of   ultrahigh energy neutrino fluxes.
The currently measured \cite{AMANDA} ultrahigh energy neutrino flux  is the
atmospheric one which results from interactions of cosmic rays with the nuclei in the atmosphere. It is rather steeply falling flux $E^{-3.7}$, however it will have the large energy tail due to the charm production which will be discussed in detail in the next section.

There are several predictions for the ultrahigh energy neutrino fluxes other than the atmospheric one, most of them differ by orders of magnitude. 
For the detailed review on  the possible ultrahigh neutrino sources see for example \cite{ML,Protheroe:rev} and the  recent constraints on the ultrahigh energy neutrino fluxes have been given in \cite{SemSigl:03}.

Here we briefly review the possible  sources of ultrahigh energy neutrinos.

One can expect a contribution to the  galactic flux resulting from interactions of cosmic rays with the interstellar medium, and the extragalactic flux which originates from the $pp$
interactions of cosmic rays with 
intracluster gas in the galaxy clusters. At the very high energies  one can also expect  cosmogenic neutrino flux coming from the interactions of the highest energy cosmic rays with the ubiquitous photons from cosmic microwave background \cite{Cosmogenic}.

An interesting possible source of the ultrahigh energy neutrinos are the Active Galactic Nuclei \cite{SalStecker,Mannheim,Protheroe:AGN,Semikoz:AGN}. These are highly luminous objects, which luminosity from the nucleus  reaches or exceeds that of our Milky Way galaxy $L_{\rm AGN}\simeq 0.1-10 L_{\rm Milky Way}$. The energy of the AGN comes from gravitational energy of a  supermassive black hole in the center of the nucleus with mass $M_{\rm BH}=10^6-10^{10} M_{\rm sun}$. The matter accrets onto a disc which surrounds a black hole, and the hot gas from inner parts of the disc can move away from it and get squeezed by the magnetic field. This leads to the emergence of highly luminous jets which can extend over the distance of $1-50 \, {\rm kpc}$.
There are different classes of AGNs like Seyfert, blasars and quasars. Most probably they are similar objects but with just a different orientation towards the Earth. The production of the neutrinos occurs through the reaction
of $p\gamma\rightarrow \Delta^+ \rightarrow \pi^+ + n$ with subsequent decay of the pion. Protons can be accelerated in the accretion disc and interact there with the  photons,
or in the jets where they can interact with the synchrotron radiation photons.
The shock acceleration has been proposed as a most plausible mechanism for the acceleration in AGNs (see for example \cite{ML}).

Gamma Ray Bursts which are characterised by the intense and short-time photon spectra can also be considered as a possible source of neutrino flux.
The detailed origin of GRB is not known, though most probably they are very massive hypernovae \cite{GRB}. The mechanism of  GRB is modelled via expanding relativistic fireball. If the protons are also shock accelerated they might
interact with the synchrotron photons to produce neutrinos via pion photo-production \cite{GRB_WB}.

Supernovae are powerful sources of the neutrinos and gamma-rays at nuclear energies. Though, there has been no evidence for the high-energy gamma ray or neutrinos fluxes from supernova SN1987A, it is possible that supernovae could be sources of the highly energetic neutrinos.

Other, more exotic mechanisms are also studied. 
In particular the annihilation of Weakly Interacting Particles (WIMPs) trapped in the cores of Sun or the Earth \cite{JKG}. WIMPs have been discussed in the context of the dark matter component of the universe. Using the predicted density for the dark matter
one can estimate the flux of neutrinos which would come from the process of WIMPs annihilation. For example within the supersymmetric scenario annihilation of neutralinos has been proposed as a source for the flux of muon neutrinos \cite{JKG}. In the supersymmetric extension of the standard model with conserved R-parity neutralino is the lightest SUSY particle-LSP and it is stable. Neutralinos could be therefore attractive candidates for the dark matter WIMPs. If their density is high enough they could annihilate producing ultrahigh energy neutrino flux in the final state \cite{JKG} (for recent calculation see \cite{Halzen}).

Another speculative source are so called top-down models \cite{Sigl:TD,Sigl:old} which assume the existence of the very heavy dark matter particles \cite{EGS,BKV} - Wimpzillas. 
In various GUT models these heavy particles occur naturally with a mass
around the GUT scale $m_X \simeq 2 \times 10^{16} {\rm GeV}$.
If their lifetime is long enough they would contribute to the cosmic and neutrino flux at very high energies. In some models very heavy particles are produced
continuously from the emission of the topological defects \cite{TDV,BHS,BV} that are relics of the cosmological phase transitions in the early universe.
For example, during the reheating phase in the early universe the ultraheavy relics can be produced quite copiously and they could be the candidates for the dark matter in the universe. These supermassive particles  with the relatively long lifetime
could be concentrated in the galactic halo \cite{Sarkar}. This is very interesting possibility since it could explain the possible excess of the data \cite{AGASA} beyond the GZK \cite{GZK} cutoff:
the products of the decays of these wimpzillas would have energies exceeding $10^{19} {\rm eV}$. 
This in consequence would also lead to the enhancement neutrino flux at the ultrahigh energies.

Finally, we stress that  an important general upper bound on the
ultrahigh energy neutrino flux has been derived \cite{WB}  from the known cosmic ray flux.
This limit is applicable to the AGNs and GRBs and could only be exceeded through the exotic scenario (like WIMPs annihillation or the top-down model )
or if the production source for the cosmic rays is  optically thick to proton-photon pion production or proton-nucleon interactions (so called ``neutrino only factories'').

\section{Atmospheric flux and prompt neutrino production}
In this section we study in detail the production of the high-energy neutrinos
 in the atmosphere. This neutrino flux constitutes an important background to the flux of cosmic neutrinos discussed in the previous section.
The  atmospheric neutrino flux is produced by the interaction of cosmic rays with the nuclei in the atmosphere \cite{Gaisser}. The neutrinos mainly come from the decays of pions and kaons. At energies of around $ 100 \; {\rm TeV}$ the decay length of these mesons is so large, that they  interact before  decaying and  loose substantial amount of energy. This effect leads to the steeply falling spectrum of atmospheric neutrinos \cite{Gaisser,Volkova}
$$
\frac{d^2 \Phi}{dE_{\nu} d\Omega} \; = \; N_{\nu} E_{\nu}^{-3.7} \; ({\rm cm^2 \; s \; sr \; GeV})^{-1} \; .
$$
At higher  energies $E_{\nu} > 100 \; {\rm TeV}$ the atmospheric neutrino flux will be dominated by the prompt neutrino production originating from charm  and also beauty hadron decays. These hadrons ($D_0,D^{+(-)},D_s^{+}$ mesons, barion $\Lambda_c$ and their $B$ counterparts) are characterised by a  very short lifetime $\sim 10^{-12} s$, so up to energies $10^9 {\rm GeV}$ they will not interact before decaying and the flat flux of the neutrinos will be produced which can extend up to these energies. This prompt neutrino flux at high energies
is a background  for the sought after cosmic neutrinos and  its precise evaluation is essential.
\subsection{$c\bar{c}$ production cross section}
There are several QCD-motivated calculations of the prompt neutrino flux \cite{TIG,GGV,PRS}. Below, we present a particular calculation which includes  small
 $x$ resummation effects as well as estimate of the gluon shadowing \cite{MRS}.

The cross section for the charm production can be evaluated within perturbative QCD framework. The high energy regime of the production process $pp\rightarrow  c\bar{c}+X$ probed by the cosmic rays is  however 
not accessible in the present colliders and reliable extrapolations of these predictions are therefore necessary. 
\begin{figure}[tbp]
\centering\includegraphics[width=0.5\textwidth]{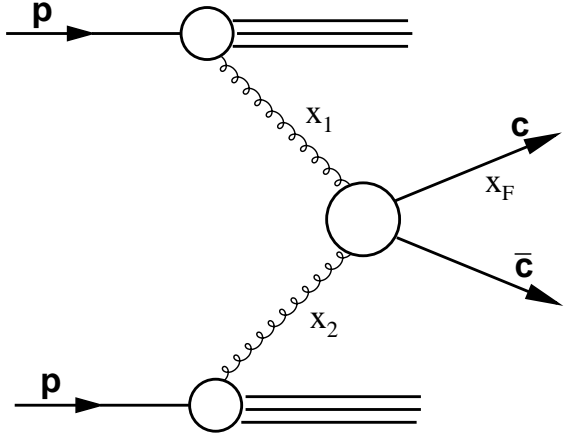}
\caption{QCD diagram for the prompt charm production in the proton-proton collisions.}
\label{fig:diagram}
\end{figure}

In Fig.~\ref{fig:diagram} we present the diagram for the  charm production in the collision of two protons. $x_F = E_c/E$ is the Feynman variable characterising the fraction of the energy of the incident cosmic ray proton carried by the produced charm quark. 
The cross section for the production of the $c\bar{c}$ pair in the proton-proton collision reads
\begin{equation}
\frac{d\sigma^{pp\rightarrow c+X}}{dx_F} \ =\ \int\,dx_1\,dx_2\,dz\
g(x_1,\mu_F^2)\:\frac{d\bar\sigma_{gg\rightarrow \bar{c}c}}{dz}\;g(x_2,\mu_F^2)\,\delta(zx_1-x_F),
\label{eq:xsection}
\end{equation}
where $\frac{d\bar{\sigma}_{gg\rightarrow \bar{c}c}}{dz}$ is the partonic cross section for the process $gg\rightarrow \bar{c}c$ and  $g(x,\mu^2)$ is the gluon density in the proton evaluated at the momentum fraction $x$ and the scale $\mu^2$. In the  domain of high energies that we are interested, the typical momentum fractions are of the order $x_1 \simeq x_F \sim 0.2-0.1$ and consequently $x_2 \simeq M_{c\bar{c}}^2/2 x_F s\sim 10^{-4}-10^{-9}$ where $M_{c\bar{c}}^2$ is the mass of the $c\bar{c}$ system. 
In general the higher the energy the lower the momentum fraction $x_2$ in question.
It means that in this process the gluon density $g(x_2,\mu^2)$ in Eq.~(\ref{eq:xsection}) is probed in the regime of  $x_2$ which is outside the range of any present accelerator data. 
We consider three approaches for an extrapolation into the low $x$ region:
\begin{itemize}
\item extrapolation of the standard DGLAP evolution
\item unifed BFKL/DGLAP evolution equations
\item inclusion of saturation effects, Golec-Biernat and Wuesthoff model
\end{itemize}

We briefly describe these calculations below.
The most straightforward approach is to use the standard gluon distribution obtained within the DGLAP(Dokshitzer-Gribov-Lipatov-Altarelli-Parisi) framework from the global fit \cite{MRST} and extend it to lower values of $x$ using the double leading-log approximation
\begin{multline}
g(x,Q^2)\ \ =\ \
g(x=10^{-5},Q^2)\,\exp\left(\sqrt{\frac{16N_C}{b}\ln\frac{\alpha_S(Q)}{\alpha_S(Q_0)}\ln\frac{x}{x_0}}
\;-\; \right. \nonumber \\
\left. - \; \sqrt{\frac{16N_C}{b}\ln\frac{\alpha_S(Q)}{\alpha_S(Q_0)}\ln\frac{10^{-5}}{x_0}}\right) \; ,
\label{eq:dlla}
\end{multline}
where $n_f=4$ and $b=25/3$. The running leading-order (LO) coupling is defined as 
$$\alpha_S(Q) =
\frac{4\pi}{b\log(Q^2/\Lambda^2_{\rm QCD})} \, .$$
 This approach correctly incorporates the leading $\alpha_s \ln Q^2 \ln 1/x$ terms.
In this calculation (called MRST on the plots) we have used the LO formulae for the partonic cross section and multiplied them by the factor $K=2.3$ which accounts for the next-to-leading (NLO) corrections.

 Another approach is to use the BFKL(Balitsky-Fadin-Kuraev-Lipatov) \cite{BFKL} framework which resums the leading terms $\alpha_s \ln 1/x$. This  approach is probably more appropriate since in the process considered, $Q^2 \simeq m_c^2$ is fixed and not too large whereas the values of  Bjorken $x$  probed are very small. We have chosen to use the gluon distribution function obtained from the solution of unifed BFKL/DGLAP evolution equations \cite{KMS1}. This approach (which we call KMS on the plots) has the advantage of treating both the BFKL and DGLAP evolution schemes on equal footing and it also resums
the major part of the NLO corrections to the BFKL equation via the so-called kinematical constraint \cite{KC}.
In this framework one is using the high energy  factorisation theorem \cite{CCH} together with the  unintegrated gluon distribution function $f(x,k^2)$
which is related to the integrated distribution $xg(x,\mu^2)$
$$
xg(x,\mu^2) = \int^{\mu^2} \frac{dk^2}{k^2} f(x,k^2) \ .
$$
Here $k^2$ is the virtuality of the gluon.
More precisely for the evaluation of the cross section one has to use the off-shell matrix element for $gg \rightarrow c\bar{c}$ convoluted with functions $f(x,k^2)$ 
\be
\frac{d\sigma^{pp\rightarrow c+X}}{dx_F} \ = \
\int dx_1 \, dx_2 \, dz \,  dk_1^2 \,  dk_2^2 \, f(x_1,k_1^2) \, \frac{d\bar{\sigma}^{\rm off-shell}}{dz} \, f(x_2,k_2^2) \delta(z x_1-x_F) \;,
\label{eq:offshell}
\ee
where $k_1^2$ and $k_2^2$ are the virtualities of the two incoming gluons
and $\frac{d\bar{\sigma}^{\rm off-shell}}{dz}$ is an off-shell matrix element for 
the $gg\rightarrow c\bar{c}$ process.
This approach is more appropriate at small $x$ since it accounts for the resummation of the $\alpha_s \ln 1/x$ terms in the splitting $gg\rightarrow c\bar{c}$. 

In the third calculation (which we refer as to GBW) we have included the effects which might come from the 
parton saturation.  In the region of high energies, where the parton densities  become high, the  gluon recombination effects can become important. These effects lead to the slower increase of the parton density with energy
and in consequence to the damping of the cross section.
This is called perturbative parton saturation. For a review on various approaches to this phenomenon,  see \cite{IV}.
In our calculation we have used very successful model by  Golec-Biernat and Wuesthoff \cite{GBW} which is formulated within the dipole picture of high energy scattering. This model has been successfully applied to the description of the inclusive and diffractive data on deep inelastic scattering at HERA.
Let us briefly sketch the philosophy of this approach which was originally formulated for the deep inelastic process of the scattering of electron on a nucleon.
In the dipole picture the virtual photon fluctuates into a $q\bar{q}$ dipole pair and then interacts with the target. These two processes are well separated in time.
The total cross section for the $\gamma^* p$ scattering can be expressed in  the following form
\begin{equation}
\sigma_{T,L}(x,Q^2) \; = \; \sum_q \int d^2 r \, dz \, |\Psi^q_{T,L}(r,z,Q^2)|^2 \; \sigma^{q\bar{q}}(r,x) \; ,
\label{eq:dipole}
\end{equation}
where $\Psi^q_{T,L}$ is the virtual photon wave function ($T$ stands for transversely and $L$ for longitudinally polarized virtual photons) and $\sigma^{q\bar{q}}(r,x)$ is the $q\bar{q}$ dipole  cross section which contains the information about the interaction between the $q\bar{q}$ dipole and the proton.

\begin{figure}[tbp]
\centering\includegraphics[width=0.7\textwidth]{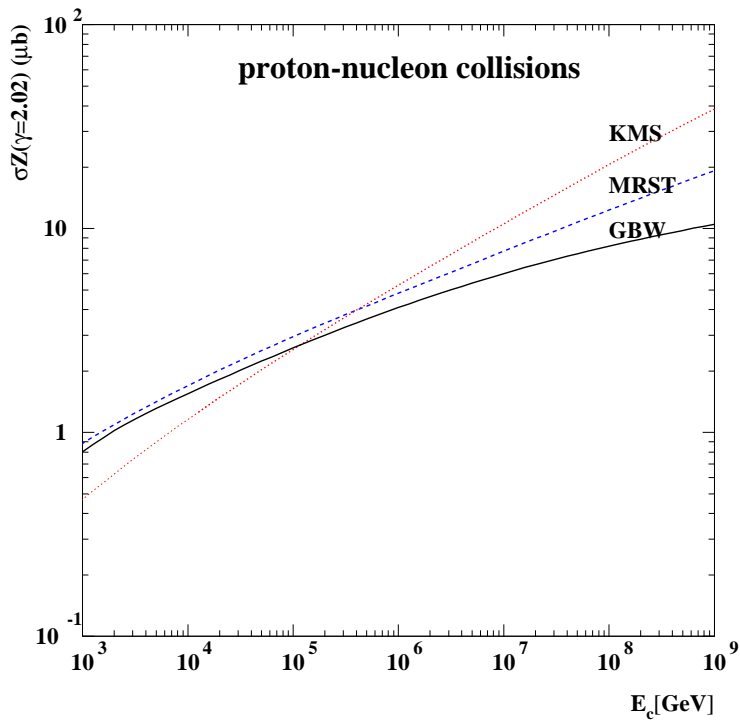}
\caption{The $Z$ moment of charm production in proton-nucleus collisions
plotted as a function of energy $E_c$ of the produced charm quark.
Solid line - GBW saturation model, dashed line - standard DGLAP calculation based on MRST partons, dotted line - KMS prediction based on the unifed BFKL/DGLAP evolution equations. The $Z$ moment is shown for illustration. In the  calculation of the neutrino fluxes the differential cross section is convoluted with the primary cosmic ray flux.}
\label{fig:sigpronuc}
\end{figure}

The $r$ is the dipole pair transverse size, and $z$ is the longitudinal momentum fraction of the photon carried by the quark.
The photon wave functions $\Psi_{T,L}^q$ are well known in leading order. In the GBW saturation model \cite{GBW}
the dipole cross section has been parametrised in the following form
\begin{equation}
\sigma^{q\bar{q}}(r,x) \; = \; \sigma_0 \left[ 1 - \exp(-\frac{r^2 Q_s^2(x)}{4}) \right] \; ,
\label{eq:gbwdipole}
\end{equation}
with the saturation scale 
\begin{equation}
Q_s(x) \; = \; Q_0 \left(\frac{x}{x_0}\right)^{-\lambda/2} \; ,
\label{eq:satscale}
\end{equation}
which governs the $x$ (or energy) behaviour.
The parameters of the GBW model are $Q_0 = 1 \, {\rm GeV}$, $x_0=3\times10^{-4}$, $\lambda=0.28$, $\sigma_0=23 \, {\rm mb}$ and are chosen to give the best description of the experimental data at HERA collider.

This model has the  property that for small dipole sizes $r \ll 1/Q_s(x)$
the cross section is also small and behaves proportional to $r^2$.
For large dipole sizes $r \gg 1/Q_s(x)$ the dipole cross section $\sigma^{q\bar{q}}$ does not grow anymore, but rather saturates to the constant value given by parameter $\sigma_0$.
The transition scale $Q_s(x)$ between the linear and saturated regimes called saturation scale grows with energy like a power (\ref{eq:satscale}). The larger the energy the smaller the dipole which will saturate.  This model has been extended from $\gamma^* g \rightarrow q\bar{q}$ to describe the $gg\rightarrow c\bar{c}$ process as well
and applied  to the prompt charm production \cite{MRS}.
Let us finally note, that GBW model contains much  more than the perturbative saturation. This model  is valid for all values of $Q^2$ and the dipole cross section extends to arbitrary large dipole sizes, therefore GBW parametrisation necessarily embodies a lot of soft or nonperturbative components.

In Fig.~\ref{fig:sigpronuc} we have summarised the results for the charm production cross section based on the three above-mentioned calculations. We have plotted a moment
$$
\sigma Z_c \; \equiv \; \int \frac{d\sigma^{pp\rightarrow c+X}}{dx} \, x^{2.02} \, dx \; .
$$
he calculation labelled MRST is the standard DGLAP approximation, KMS includes additional small $x$ dynamics through the unifed BFKL/DGLAP evolution  whereas GBW accounts for the gluon saturation.
We see that,  the calculation based on unifed DGLAP/BFKL equation
gives the highest prediction for the  energies $> 10^6 \, {\rm GeV}$, whereas the 
calculation which contains the gluon saturation tends to have slowest increase with energy. This is because the parton distributions obtained from solution to the unifed DGLAP/BFKL evolution equations exhibit a stronger growth with energy than the ones from standard calculation within the DGLAP approach. 
The difference between the KMS and MRST calculations at lowest energies can be explained by the fact that the latter has been multiplied uniformly by the next-to-leading  K factor ($\simeq 2.3$), whereas the KMS calculation uses the off-shell matrix element which in principle contains subleading corrections, important at high energies.
On the other hand, in the low energy regime the gluon distribution is probed at rather large values of $x\sim 0.1$ and so the $\ln 1/x$ resummation in matrix element in KMS calculation is not effective. The KMS prediction should be in principle multiplied by the energy dependent K factor which would vanish in the high energy region.

\subsection{Calculation of the prompt $\nu_{\mu}$ and $\nu_{\tau}$ flux.}
In order to calculate the final neutrino flux one has to follow the sequence
of the charm production, charm quark fragmentation and charm meson decay
into neutrinos. Additionally the attenuation of the cosmic rays has to be taken into account as well as possibility of the interaction of the charmed mesons.
This sequence of processes has been followed by use of transport equations
(see for example \cite{Gaisser} )
which are differential equations  in  depth $X$ of atmosphere traversed by the particle
$$
X \; = \; \int_h^{\infty}\, \rho(h') \, dh' \; ,
$$
where $\rho(h)$ is  the density of the atmosphere as measured at height $h$.
The density profile has been taken to be exponential $\rho(h) = \rho_0 \, \exp(-h/h_0)$ with $\rho_0 = 1.6\times 10^{-2}\,{\rm g/cm^3}$ and $h_0=6.4 \, {\rm km}$.
\begin{figure}[tbp]
\centering\includegraphics[width=0.7\textwidth]{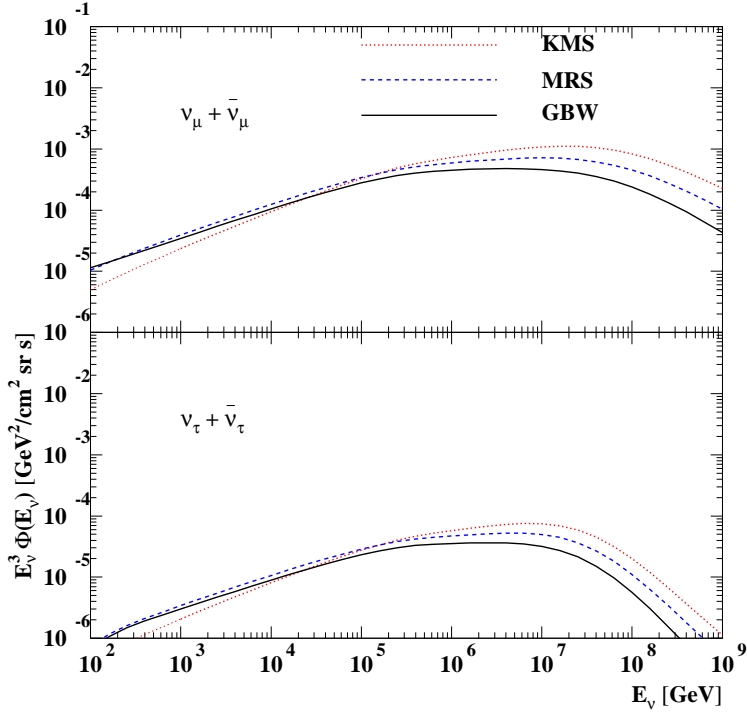}
\caption{The prompt
$\nu_\mu+\bar{\nu}_\mu$ and $\nu_\tau+\bar{\nu}_\tau$ fluxes,
arising from $\cc$ production, fragmentation and decay, obtained
using the three different extrapolations of the gluon to very
small $x$.}
\label{fig:fluxkmg}
\end{figure}
The transport equations which determine the fluxes have the following form 
\begin{multline}
\frac{\partial\phi_N(E,X)}{\partial X}\ =\
-\frac{1}{\Lambda_N(E)}\phi_N(E,X)\, ,  \\
\phi_c(E,X) \, =\, \int_E^\infty
dE'dx_c\phi_N(E',X)\frac{n_A}{\rho}\frac{d\sigma}{dx_c}^{p\ra c}
\delta(E-x_cE')\, ,  \\
\frac{\partial \phi_i(E,X)}{\partial X}\ =\
-\frac{1}{\Lambda_i(E,X)}\phi_i(E,X)+\int_E^\infty
dE'dx\phi_c(E',X)\frac{dn}{dx}^{c\ra i} \delta(E-xE') \, ,  \\
\frac{\partial\phi_l(E,X)}{\partial X}\ =\
\sum_i\int_E^\infty dE'dx\phi_i(E',X)\frac{1}{\lambda_i^{\rm
dec}(E',X)}B(i\ra l)\frac{dn}{dx}^{i\ra l} \delta(E-xE') \, , \\
\label{eq:fluxes}
\end{multline}
\begin{figure}[tbp]
\centering\includegraphics[width=0.7\textwidth]{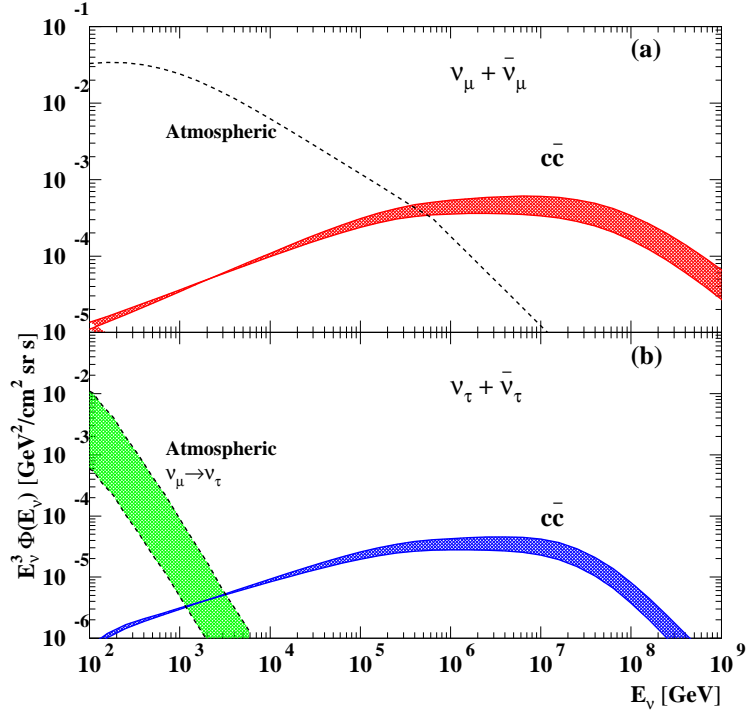}
\caption{The prompt
(a)~$\nu_\mu+\bar{\nu}_\mu$ and (b)~$\nu_\tau+\bar{\nu}_\tau$
fluxes calculated using the charm production cross sections
obtained from GBW model. Also shown are the
conventional muon and tau neutrino atmospheric fluxes, where the
latter originates, via neutrino mixing transitions, from the
former. There is also a contribution to the prompt $\nu_\tau +
\bar{\nu}_\tau$ flux from beauty production, which is not included
here, but is shown in Fig.5.}
\label{fig:fluxgbw}
\end{figure}
Here, $\phi_N(E,X)$ is the cosmic ray flux at the depth height $X$, with $\phi_N(E,0)$ being the
initial flux (primary cosmic ray flux). $\phi_c(E,X)$ is the $c$  quark flux, $\phi_i(E,X)$ is the charmed hadron flux (where $i=D^0,D^{\pm},\bar{D}^0,D_s^{\pm},\Lambda_c$) and $\phi_l(E,X)$ is the lepton flux.
The nucleon attenuation length is defined as
$$
\label{eq:Lambda_N} \Lambda_N(E) \equiv
\frac{\lambda_N(E)}{1-Z_{NN}}\, ,
$$
where $Z_{NN}$ is the spectrum-weighted moment for the nucleon regeneration and $\lambda_N$ is the interaction thickness.

\begin{figure}[tbp]
\centering\includegraphics[width=0.7\textwidth]{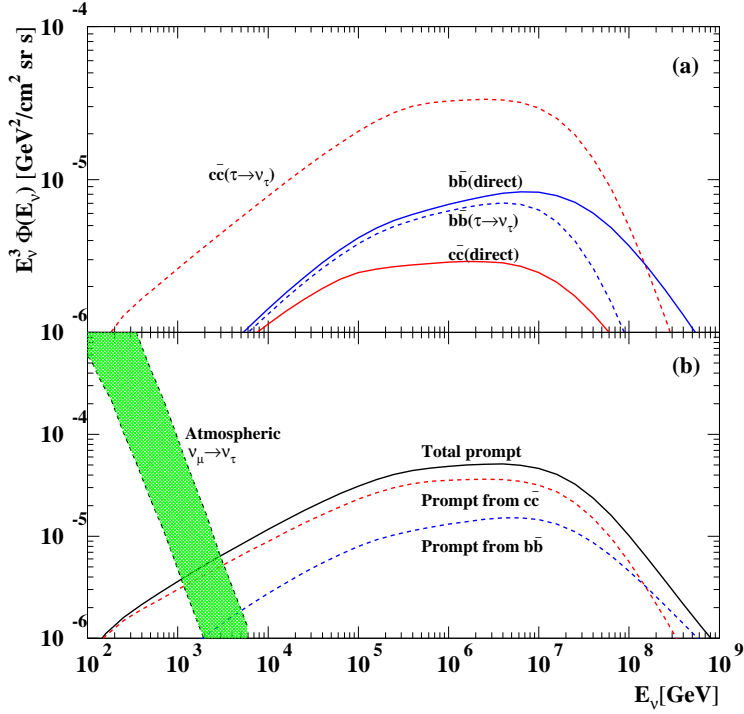}
\caption{The prompt
$\nu_\tau+\bar \nu_\tau$ fluxes originating from $\cc$ and $\bb$
production and decay, which respectively arise from the
$D_s\ra\tau\nu_\tau$ decay, and from the $B^\pm$, $B^0$, $B_s$ and
$\Lambda_b$ semileptonic $\tau\nu_\tau$ decay modes. The upper
plot shows the breakdown into the direct $\nu_\tau$ contribution
(continuous curves) and the indirect $\tau\ra\nu_\tau$
contribution (dashed curves). The lower plot shows the total
prompt $\nu_\tau + \bar{\nu}_\tau$ flux, together with its
components of $\cc$ and $\bb$ origin. Also shown is the non-prompt
$\nu_\tau + \bar{\nu}_\tau$ flux arising from $\nu_\mu \ra
\nu_\tau$ oscillations from the conventional atmospheric $\nu_\mu$
flux.}
\label{fig:fluxb}
\end{figure}

 The attenuation length for charmed hadrons has two 
terms: decay length and interaction length
$$
\label{eq:Lambda_i} \frac{1}{\Lambda_i}\ =\
\frac{1}{\lambda_i^{\rm dec}} + \frac{1-Z_{\cc}}{\lambda_i} \; ,
$$
where $\lambda_i^{\rm dec}$ is the decay length and $\lambda_i/(1-Z_{\cc})$
is the attenuation length for the charmed hadron with $Z_{\cc}$ 
being the charm regeneration factor.
The branching fraction of the decay of the charmed hadron into leptons
is given by $B(i \ra l)$. The differential cross section $\frac{d\sigma_{p\ra c}}{dx_c}$ is given by the calculation presented in the previous section. The distribution $\frac{dn^{c\ra i}}{dx}$ describes the fragmentation of the charm quark into the hadron and we have taken it to be proportional to  $\delta(x_D-0.75 x_c)$
rather than using complete fragmentation function. Finally the lepton distribution from the hadron decay $\frac{dn^{i \ra l}}{dx}$ has been taken from PYTHIA
Monte Carlo \cite{PYTHIA}.
 
In Fig.~\ref{fig:fluxkmg} we present the predictions for the $\nu_{\mu}$
and $\nu_{\tau}$ fluxes coming from the solution to the transport equations (\ref{eq:fluxes}) based on three models of the gluon extrapolation in $\frac{d\sigma^{p\rightarrow c}}{dx_c}$ to small values of  $x$. The KMS prediction is highest at high energies due to the steepest increase of the gluon density caused by the resummation of the $\alpha_s \ln(1/x)$ terms. On the other hand the GBW saturation model gives lowest predictions, due to the gluon saturation effects
embodied in the model. To be more precise we have considered several scenarios for the saturation(for details see \cite{MRS}), see Fig.~\ref{fig:fluxgbw} where the band in both plots shows the uncertainty  of the model due to the different assumptions (like the strength of the 3-pomeron coupling). The prediction for the $\nu_{\mu}$ flux is compared with the standard atmospheric flux  coming from the decay of lighter mesons: $\pi$ and $K$. The prompt $\nu_{\tau}$  flux is practically the only source for the atmospheric neutrinos at energies $E>10 \; {\rm TeV}$. Additionally we have a  $\nu_{\tau}$ component  which comes from the oscillations $\nu_{\mu} \ra \nu_{\tau}$, dominant at lower energies $E < 1 \; {\rm TeV}$.

Apart from charm decay it is also possible to have neutrino fluxes from the beauty hadron decays. For the $\nu_{\mu},\nu_{e}$ fluxes these processes
are negligible, but for the $\nu_{\tau}$ the  $B$ decays are an important
source of the flux. This happens because due to larger mass of $B$ hadrons with respect to charmed hadrons  more decay channels to $\nu_{\tau}$ are opened (in the case of charm only $D_s$ gives contribution to $\nu_{\tau}$ ) and this compensates for the lower value of the production cross section. In Fig.~\ref{fig:fluxb} we show the predictions for the $\nu_{\tau}$ flux from $\bb$ decays as compared to $\cc$ decays and we see that at energies around $10^6 {\rm GeV}$ the $\bb$ component constitutes about  $30 \%$ of the total $\nu_{\tau}$ flux, and at higher energies this proportion is even bigger.

\section{Neutrino cross sections at high energies}

Another topic that relates QCD at small $x$ and the ultrahigh energy neutrino
physics is the problem of the neutrino interactions with the encountered matter on their way to the detector. The highly energetic neutrinos will interact
with electrons and nucleons in the Earth and the detailed knowledge about the cross sections is essential for the knowledge of the rate of muons to be observed in the detector. There are various types of interactions of neutrinos.
The interactions with electrons of the muon and electron neutrinos can be listed as follows
$$\nu_{\mu} e \rightarrow \nu_{\mu} e,\bar{\nu}_{\mu} e \rightarrow \bar{\nu}_{\mu} e,
\nu_{\mu} e \rightarrow \mu \nu_e,
\nu_e e \rightarrow \nu_e e \; , $$
and the interactions of the electron antineutrinos
$$\bar{\nu}_e e\rightarrow \bar{\nu}_e e,
\bar{\nu}_e e\rightarrow \bar{\nu}_{\mu} \mu,
\bar{\nu}_e e \rightarrow {\rm hadrons} \; . $$
The first group of interactions is characterised by the constant cross section at high energies. On the other hand the interactions $\bar{\nu}_e e \rightarrow {\rm anything}$ have a characteristic sharp peak coming from a Glashow resonance at $E_{\nu} \simeq 6\,{\rm PeV}$ due to the $W^{-}$ production. 

This window of the enhanced electron antineutrino  interactions
together with the so-called double bang events in the case of tau neutrinos \cite{DOUBLEBANG}
can be used to disentangle the flavour composition of the incoming 
cosmic neutrino flux in the neutrino observatories.

\begin{figure}[tbp]
\centering\includegraphics[width=0.8\textwidth]{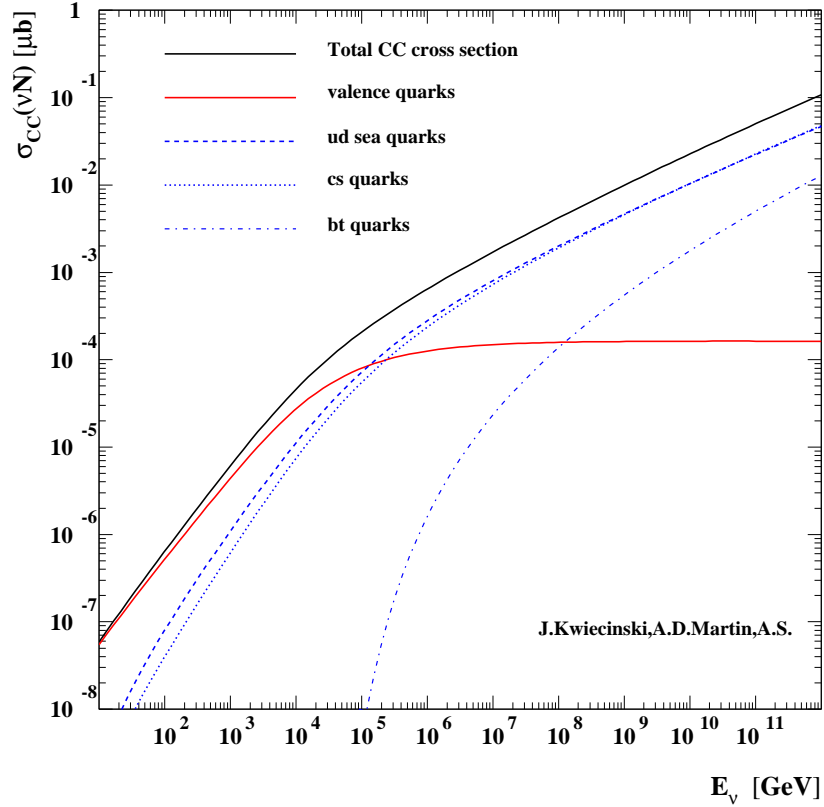}
\caption{The total $\nu N$ charged current cross section and its decomposition into components of different origin (valence,sea quarks) as a function of the laboratory neutrino energy $E_{\nu}$.}
\label{fig:epow}
\end{figure}

The interactions with nucleons proceed through the charged and neutral current deep inelastic process $\nu N \stackrel{CC}{\rightarrow} l N,\nu N \stackrel{NC}{\rightarrow} \nu N$.
The cross section for the charged current interaction, $\nu N \stackrel{CC}{\rightarrow} l N$ scattering
on the isoscalar target $N=\frac{n+p}{2}$
has the following form
\begin{equation}
\frac{d^2\sigma^{CC}}{dx dy} \, = \, \frac{2 G_F^2 M_N E_{\nu}}{\pi}
\left(\frac{M_W^2}{Q^2+M_W^2} \right)^2 \cdot \left[xq(x,Q^2)+x\bar{q}(q,Q^2)(1-y)^2 \right] \; ,
\end{equation}
where $M_N$ mass of the target nucleon, $Q^2$ is the exchanged vector boson four momentum transfer,
$x=Q^2/(2M\nu)$ Bjorken variable, $y=\nu/E$ inelasticity (where $\nu=E_{\nu}-E_l$
 is energy loss in the lab frame), $G_F$ Fermi coupling constant and  $M_W$ mass of the $W$ boson.
$xq(x,Q^2)$ and $x\bar{q}(x,Q^2)$ are the quark (antiquark) momentum distributions evaluated at the momentum fraction $x$ and at scale $Q^2$.

\begin{figure}[tbp]
\centering\includegraphics[width=0.8\textwidth]{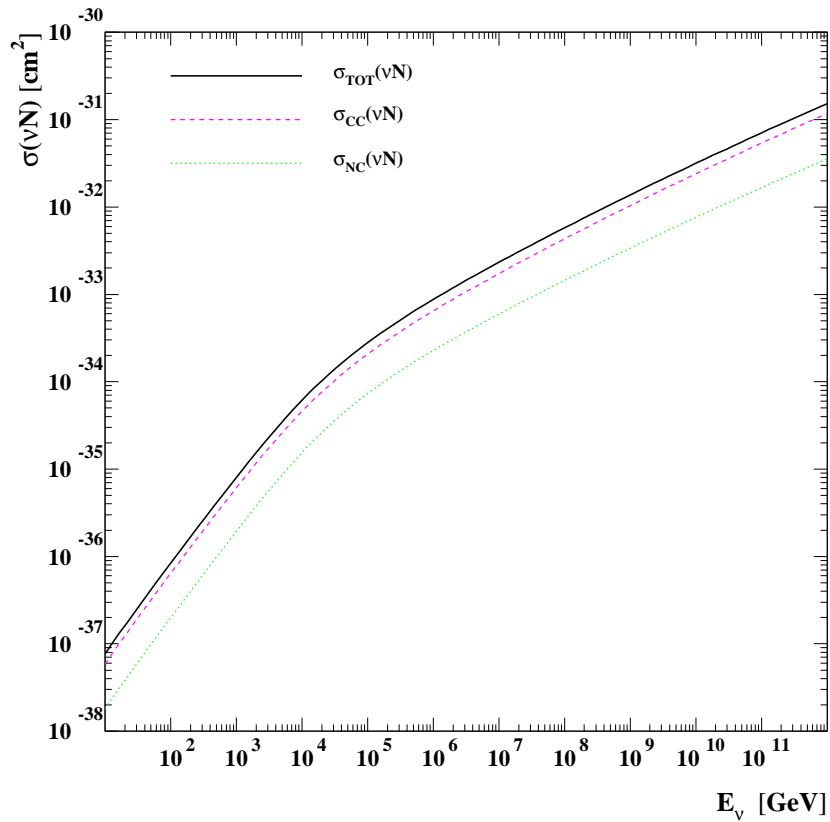}
\caption{The total $\nu N$  cross section and its decomposition into charged current and neutral current components as a function of the laboratory neutrino energy $E_{\nu}$.}
\label{fig:ccnc}
\end{figure} 

The characteristic feature of this deep inelastic cross section
is the fact that the parton densities grow with decreasing values of $x$ as 
$xq(x,Q^2) \sim x^{-\lambda}$ as we know from the measurements of the proton structure function at HERA collider \cite{HERA}. This results in the strong growth of the total $\nu N$
cross section with energy
\begin{equation}
\sigma^{tot}(E) \, = \, \int dx dy \frac{d^2 \sigma}{dx dy} \sim E^{\lambda} \; .
\label{eq:epow}
\end{equation}
The behaviour of the charged current cross section with energy is illustrated in Fig.~\ref{fig:epow} where the QCD calculation \cite{KMS2} of the partonic densities has been done within the framework of the unifed BFKL/DGLAP equations \cite{KMS1}.

We show various components to the cross section. First thing to note is the fact
that above $10^5 {\rm GeV}$ sea quarks($ud$,$cs$) start to dominate over the valence distribution.
The valence quark contribution grows linearly with energy for $E<10^5 {\rm GeV}$ 
due to the opening of the phase space, and then flattens out because 
the factor $M_W^2/(Q^2+M_W^2)$ provides a cutoff and because of the fact that the number of the valence quarks is finite.
On the other hand the density of sea quarks grows as a power with $x$ which is translated into
a power growth with energy as in (\ref{eq:epow}).
Threshold effects in the production of heavy quarks ( $bt$ ) are also visible at lower energies.
In Fig.~\ref{fig:ccnc} we show the total cross section together with the charged and  neutral current components. The neutral current cross section constitutes about  $30 \%$ of the total cross section at high energy, the fraction being 
determined by the ratio of combinations of the chiral couplings.

\begin{figure}[tbp]
\centering\includegraphics[width=0.8\textwidth]{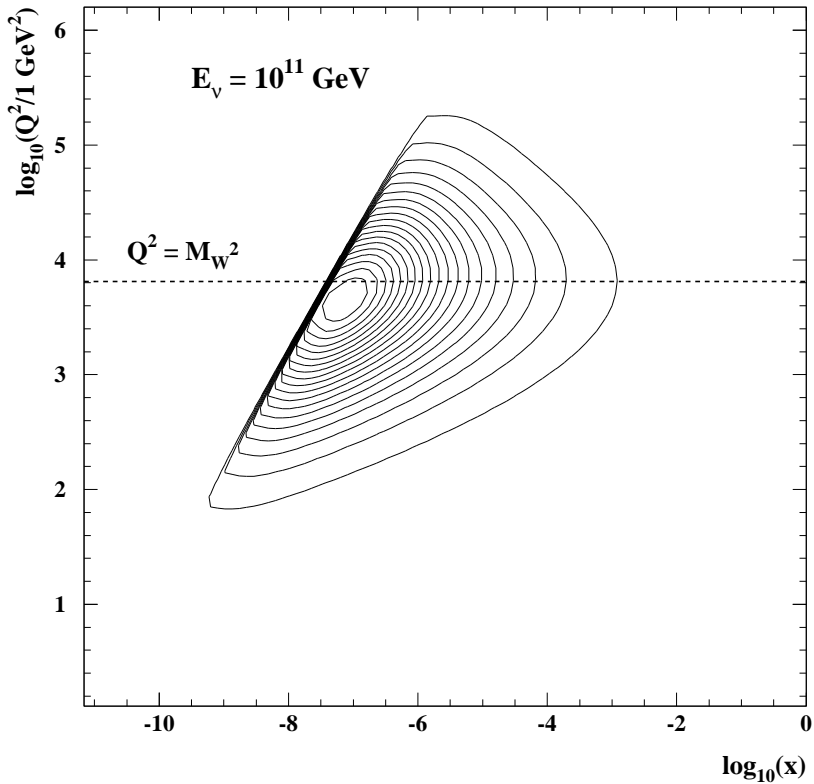}
\caption{The contour plot showing the $(x,Q^2)$ domain of the dominant contribution to the $\frac{d^2\sigma}{d\ln(1/x) d\ln(Q^2/\Lambda^2)}$ for the total $\nu N$ interaction for neutrino laboratory energy $E=10^{11} \, {\rm GeV}$. The 20 contours are such that they enclose  contributions of $5,10,15,\dots \%$ of the above differential cross section.}
\label{fig:2d}
\end{figure}

One of the important questions is to what extent are these predictions
stable. After all, one is extrapolating these calculations a long way into the new kinematical domain. There are various sets of the parton distributions which
can have  different asymptotic behaviour at very small $x$.
However, as has been investigated in \cite{KMS2} and \cite{GQRS} the parton distributions are quite well constrained by the current experimental data and so  variations
of these distributions result at most in factor $2-3$ difference in normalisation at highest energies.

To visualise the kinematic regime probed in ultrahigh energy neutrino-nucleon
interactions we have made the contour plot, see Fig.~\ref{fig:2d}, of the differential cross section  $\frac{d^2\sigma}{d\ln 1/x d\ln Q^2/\Lambda^2}$ in $(x,Q^2)$ plane. The contours
enclose the regions with different contributions to the total cross section $\sigma(E_{\nu})$.
We see that for the very high energy $E=10^{11} {\rm GeV}$ the dominant contribution
comes from the domain $Q^2 \simeq M_W^2$ and $x_{\rm min} \simeq M^2_W /(2M_N E) \sim 10^{-7}$ where $M_N$ is the nucleon mass. 

\subsection{High partonic densities and saturation}

In the limit of very high energies one might expect the shadowing (or saturation) 
corrections to appear and modify the asymptotic behaviour of the total cross section. In Ref.~ \cite{Unitarity} it was argued, on the basis of the GBW saturation model \cite{GBW},
that due to the fact that the saturation scale is typically much smaller than the average value of the momentum transfer $Q_s(x) \ll \left<Q\right> \simeq M_W$,
the actual size of the shadowing corrections is negligible, since
the contribution to the cross section from the nonlinear regime $Q < Q_s(x)$
is  very small.  A more detailed estimate has been performed \cite{KK} where the nonlinear Balitsky-Kovchegov \cite{BK} has been solved which takes into
account the saturation effects. It was shown that, even though the typical values of momentum transfer $Q^2$ are in the linear regime, the nonlinear effects can tame the growth of the total cross section and reduce the magnitude at highest energies $\sim 10^{12} {\rm GeV}$ by  a factor of $2$ or so. It was shown in  \cite{KS,IIMCL} that the partonic distribution persists to feel the nonlinearity
even for momenta higher than the saturation scale, more precisely in the window  $Q_s^4(x)/\Lambda_{QCD}^2 > Q^2 > Q_s^2(x)$. Therefore the saturation effects can become non-negligible even at highest energies and they could be enhanced by the scattering of the neutrino on the nucleus.

\section{Penetration through the Earth}

\begin{figure}[tbp]
\centering\includegraphics[width=0.8\textwidth]{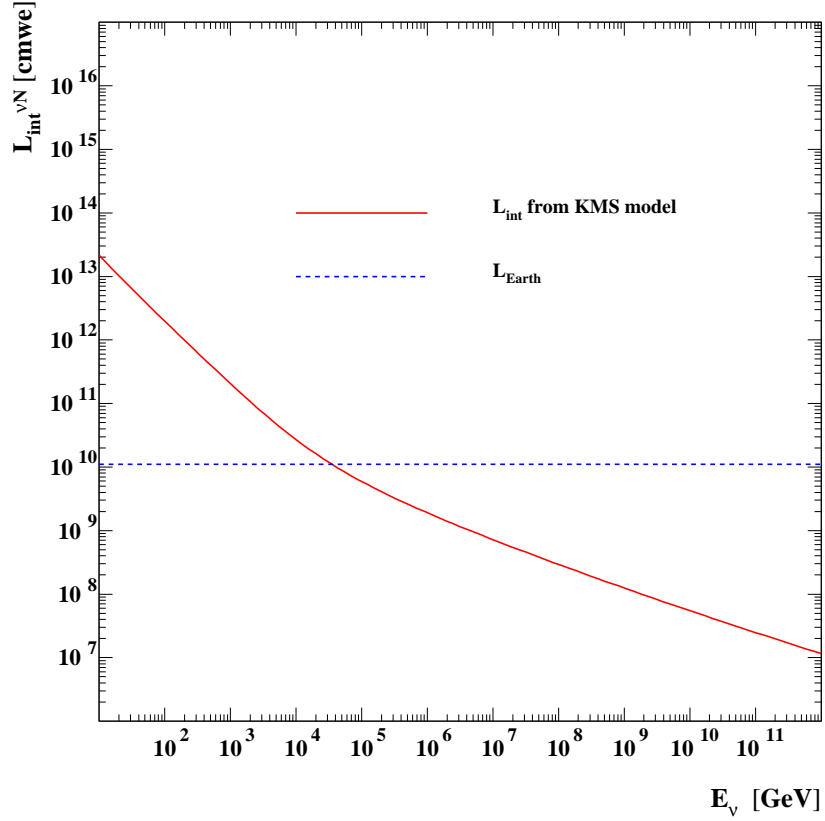}
\caption{The interaction length for the neutrino interaction with isoscalar nucleon target obtained from the KMS model. The straight line corresponds to the maximal column depth length which can be encountered by the neutrino traversing the Earth through the middle. }
\label{fig:Length}
\end{figure}

We have mentioned at the beginning that the neutrinos interact very weakly and therefore do not get disturbed during their way from the source to the detector on Earth. However, for the ultrahigh energy neutrinos the situation changes due to the increasing value of the neutrino-nucleon cross section, see Figs.~\ref{fig:epow},\ref{fig:ccnc}.
The ultrahigh energy neutrinos will undergo attenuation when penetrating
Earth \cite{GQRS}.
It is well illustrated in Fig.~\ref{fig:Length} where the interaction length
${{\cal L}_{\rm int}(E_{\nu})} = \frac{1}{\sigma^{\nu N}(E_{\nu}) N_A}$ is shown together with the maximal column depth ${\cal L}_{\rm Earth} = 1.1 \times 10^{10} {\rm cmwe}$ which is encountered by neutrino
traversing along the diameter of the Earth.  It is obvious that the interaction length becomes shorter for neutrinos with energies exceeding $40\, {\rm TeV}$. Thus Earth
will start to become opaque to highly energetic neutrinos \cite{GQRS}.

These effects can be studied using transport equations for the neutrino flux
\begin{equation}
\frac{d I(E,\tau)}{d \tau} \; = \; -\sigma_{TOT}(E) I(E,\tau) +
\int \frac{dy}{1-y} \frac{d\sigma_{NC}(E',y)}{dy} I(E',\tau) \; ,
\label{eq:Earthtrans}
\end{equation}
where the total cross section is the sum of charged and neutral current components $\sigma_{TOT}=\sigma_{NC}+\sigma_{CC}$ and where $y$ is the fractional energy loss $E' = \frac{E}{1-y}$.
$\tau = N_A X(z) = N_A \int_0^z dz' \rho(z')$ with $\rho(z)$ being the density
of the Earth along the neutrino path and $N_A$ is Avogadro number.

\begin{figure}[tbp]
\centering\includegraphics[width=0.9\textwidth]{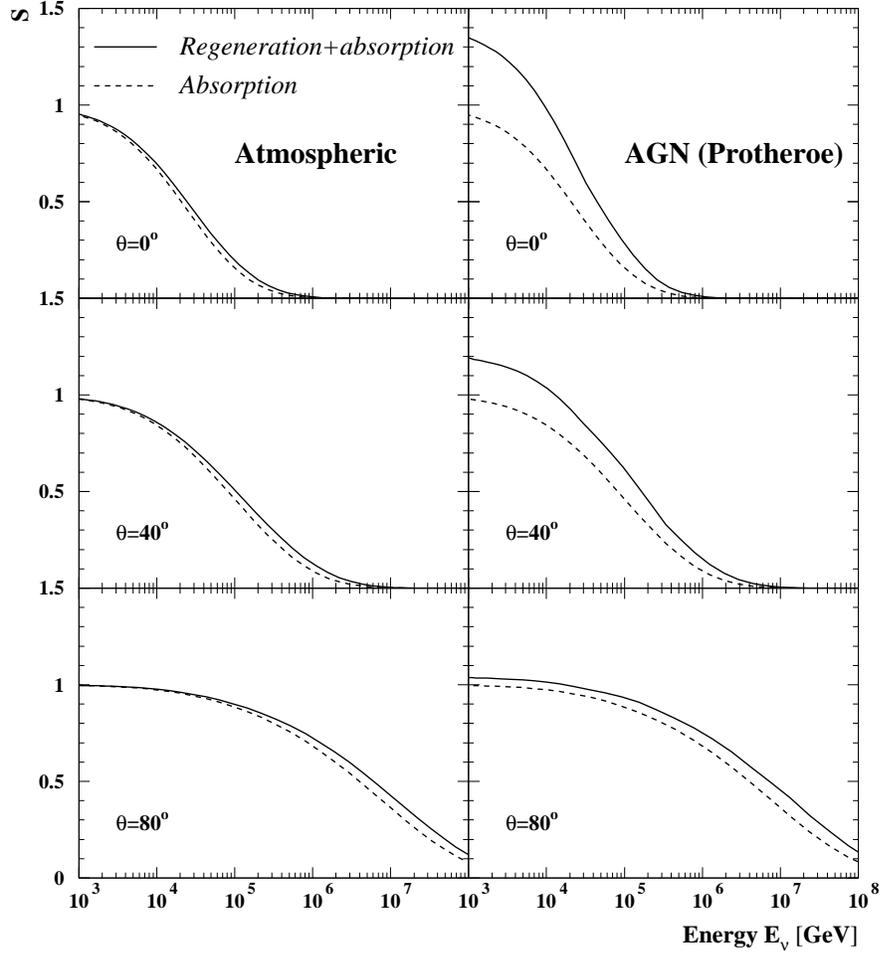}
\caption{The shadowing factor $S(E,\tau) =  \frac{I(E,\tau)}{I^0(E)}$
as a function of the neutrino energy $E_{\nu}$ for three different incident angles
$\theta = 0^o,40^o,80^o$ and two different models for the fluxes: atmospheric
(left plots) and AGNs (right plots). The solid line corresponds to the solution
of the full transport equation whereas the dashed line corresponds 
to the shadowing factor being purely absorption 
$S(E,\tau) = \exp(-\sigma^{tot}(E) \tau)$ without regeneration.}
\label{fig:Sfact}
\end{figure}

The total depth length $\tau$ depends on the incident nadir angle
which is defined as the angle between the direction of the neutrino arrival
and the normal to the Earth surface passing through the detector.
The first term in Eq.(\ref{eq:Earthtrans}) corresponds to the neutrino attenuation whereas the second term is responsible for the neutrino regeneration through the neutral current interactions at higher energies.
It is useful to study the general properties of the solution by showing the shadowing factor $S(E,\tau) = I(E,\tau)/I_0(E)$ for two different forms of initial flux, Fig.~\ref{fig:Sfact}. First is the steeply falling atmospheric \cite{Volkova} neutrino flux $I_0 \simeq c E^{-3.7}$ and second is the Active Galactic Nuclei flux \cite{Protheroe:AGN} 
which is characterised by rather flat behaviour in the interval $10^3 < E < 10^5 {\rm GeV}$.

 Three different incident angles are considered in Fig.~\ref{fig:Sfact} which correspond to different values of $X=1.1 \times 10^{10},0.45\times 10^{10},0.072\times 10^{10} \; {\rm cmwe}$ for $\theta=0^o,40^o,80^o$ correspondingly. Additionally we show the pure attenuation factor $A=\exp(-\sigma_{\rm TOT} \tau)$ which would occur, if the neutral current regeneration given by the second term in Eq.(\ref{eq:Earthtrans}) would not be present. It is apparent that for the large column depths, the attenuation at high energies is very strong, essentially only about $10\%$ of the 
initial flux survives at energies $100\, {\rm TeV}$ and the flux gets completely absorbed at yet higher energies. What is also visible is the fact
that the regeneration due to the neutral current interaction is most prominent
for flat fluxes like the one from AGN,  and rather negligible for steeply falling spectra like the atmospheric one.
In some range of energies the effect of regeneration is quite large,
in which case the shadowing factor is bigger than $1$, see Fig.~\ref{fig:Sfact}.


\begin{figure}[tbp]
\centering\includegraphics[width=0.9\textwidth]{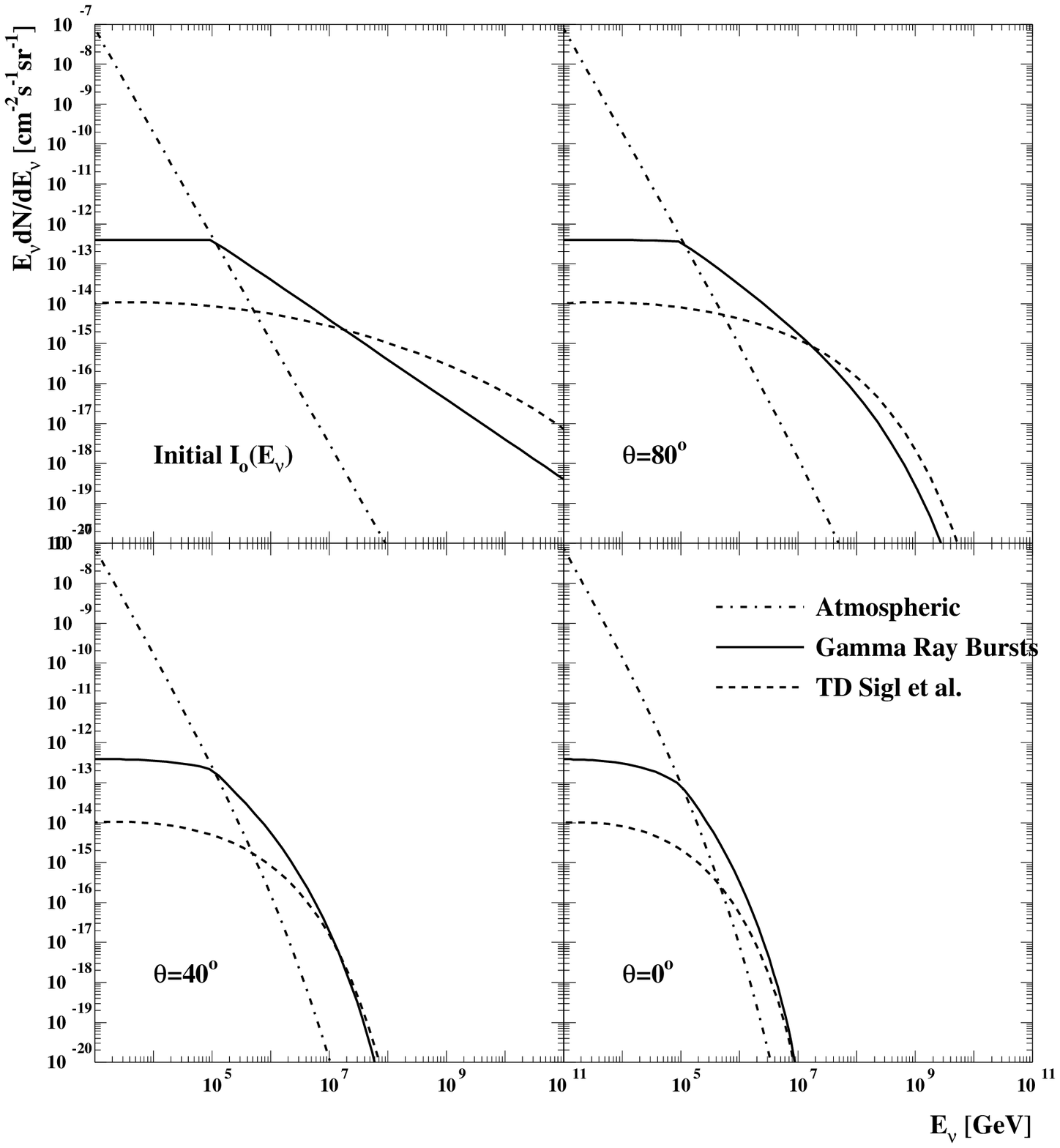}
\caption{The neutrino flux $I(E,\tau)$ as a function of the neutrino energy $E_{\nu}$ for  four different incident angles
$\theta = 0^o,40^o,80^o$ and $180^o$ (initial flux $I^o(E)$) and three different models for the fluxes: atmospheric, TD model and the GRB model.}
\label{fig:Flux}
\end{figure}

In Fig.~\ref{fig:Flux} we show the resulting neutrino flux $I(E)$ reaching the detector at different nadir angles $\theta$ as a function of energy for 
GRB model \cite{GRB_WB}, Top-Down  model \cite{Sigl:old} and the atmospheric flux.
The GRB and TD fluxes dominate over the atmospheric flux at energies above $10^5 {\rm GeV}$. However at small nadir angles we see the strong attenuation
of all fluxes at higher energies. The smaller the nadir angle the greater the shadowing.
Therefore one has to adopt different method for searching the ultrahigh energy
neutrino events in the detectors whenever $E> 1 \, {\rm PeV}$.
For smaller energies it is sufficient to look for the upgoing muon events,
which are always neutrino induced. This reduces the background from the atmospheric muons. For higher energies due to the discussed Earth shadowing 
one has too look for the horizontal or downgoing highly energetic muon events.
A detailed study on the possibility of detecting tau neutrinos using horizontal arrays has been performed in \cite{FARGION}.
\subsection{Regeneration of $\tau$ neutrinos}
It has been pointed in \cite{TAU1} and further studied in \cite{TAU2,TAU3} that the propagation of tau neutrinos 
through the Earth is quite different from the muon and electron
neutrinos. The Earth in principle never becomes opaque to the high energy tau neutrinos \cite{TAU1}. This is due to the very short lifetime of the $\tau$ lepton,
which decays quickly, producing in the final state tau neutrino still with very high energy.
This process continues until the $\nu_{\tau}$ energy is reduced to the value
at which the interaction length is equal  to the column depth 
through the Earth. The  tau neutrinos will thus
never disappear from the high energy spectrum, the flux will be only redistributed towards the lower energies with possible pile-up around the transparency energy. This is different than in the case of muons, which live substantially longer and loose their energy via interactions. When they finally decay, they produce a muon neutrino with very low energy. The effect of tau regeneration is mostly prominent for the flat fluxes, where significant enhancement can been seen around the transparency energy \cite{TAU2,TAU3}. On the other hand  for the steeply
falling fluxes the regeneration effect is negligible.

\subsection{Neutrino oscillations}

In this section we are going to study how the fluxes of neutrinos at high energies  are modified
by the oscillation mechanism. It is now  well established experimentally \cite{OSCILLATIONS} that
neutrinos undergo flavour-changing oscillations \cite{Pontecorvo}. Due to the high energy
of extragalactic neutrinos, the oscillation length is much larger than 
the diameter of the Earth so the probability that ultrahigh energy neutrinos
will oscillate on their way from the surface of the Earth to the detector
is very small. 
In fact, the oscillation length in vacuum increases linearly with energy
\be
\label{eq:lvac}
l_{v} \; =  \; \frac{4\pi E}{\Delta m_{\nu}^2} \; .
\ee
If one takes the matter effects into account \cite{MSW}, then
the effective oscillation length in matter and the effective mixing angle
become
\be
l_m \; = \; l_v \left[1+2 \frac{l_v}{l} \cos \theta_v + \frac{l_v^2}{l^2} \right]^{-1/2} \; , \; \;
\tan 2\theta_m \; = \;  \frac{\sin 2\theta_v}{\cos 2\theta_v + l_v/l} \; ,
\label{eq:msw}
\ee
where $l$ originates from the matter contribution to the oscillation length
and is given by 
\begin{equation}
l \; = \;  \frac{\sqrt{2}\pi}{G_f n_e} \; = \;  \frac{1.77 \times 10^7}{ \rho_e} {\rm m} \; \, .
\label{eq:lval}
\end{equation}
Here $n_e$ is the electron density and $\rho_e$ is the number density in units of Avogadro number$/cm^3$.
An important property of $l_m$, is the fact that it saturates at the value $l$ given by (\ref{eq:lval}). However, at sufficiently high energy $l_v/l$
is very large and the effective mixing angle $\theta_m$ tends to zero. 
This means that for large energies $E>1 \; {\rm TeV}$ one can safely neglect the 
 matter oscillation effects for the neutrinos in their passage through the Earth.

On the other hand, oscillations can change the initial neutrino flux as compared to the primary neutrino flux produced  far in the 
extragalactic source. This is of course due to very large distances  which  neutrinos have to cover on their way from the cosmic source to the Earth. It is believed, that the neutrinos emerge as a decay products of the pions which are produced 
in $p\gamma$ and $pp$ collisions.  Since $\pi^{\pm} \rightarrow \mu^{\pm} + \nu_{\mu}\rightarrow e^{\pm} + \nu_{e} + \nu_{\mu} + \bar{\nu}_{\mu}$ neutrinos are produced in ratios $1:2:0$ for $\nu_e:\nu_{\mu}:\nu_{\tau}$. More precisely the $\nu_{\tau}$ flux is produced \cite{Webber} but is  smaller than
$\nu_e$ and $\nu_{\mu}$ flux
and comes from the decay chain of heavy mesons  in the $pp$ collisions via the similar process as described in Sec.2.1 in the case of the prompt lepton  production
in the atmosphere.
Due to the very long path to the Earth the arriving neutrinos
will be completely mixed \cite{AJY} which will result in the equal numbers for all species
$1:1:1$.
Let us recall (see for example \cite{AJY}) that in the three flavour framework the mass and the flavour
eigenstates are related by means of the unitary mixing matrix $U$
\begin{displaymath}
\left[
\begin{array}{c}
\nu_e\\
\nu_{\mu} \\
\nu_{\tau}
\end{array}
\right]
\; = \;
U \left[
\begin{array}{cc}
\nu_1\\
\nu_2\\
\nu_3
\end{array}
\right] \; .
\label{eq:MNS}
\end{displaymath}
The probability of the flavour oscillation from the  neutrino
$\nu_{\alpha}$ to neutrino $\nu_{\beta}$ after travelling length $L$
is given by 
\begin{equation}
P(\nu_{\alpha} \rightarrow \nu_{\beta};L) \; = \; \delta_{\alpha\beta}-
\sum_{j\neq k} U^*_{\alpha j} U_{\beta j} U_{\alpha k} U^*_{\beta k}
(1-\exp(-i\Delta E_{jk} L)) \; .
\end{equation}
In the limit when $L \rightarrow \infty$  one gets
\begin{equation}
\label{eq:transp}
P(\nu_{\alpha} \rightarrow \nu_{\beta};L\rightarrow \infty) \; = \; \delta_{\alpha\beta} \; - \; \sum_{j\neq k} U^*_{\alpha j} U_{\beta j} U_{\alpha k} U^*_{\beta k} \; = \; \sum_j |U_{\alpha j}|^2 |U_{\beta j}|^2 \; .
\end{equation}
The cosmic neutrino flux  $F(\nu_\alpha)$ after travelling very long distance can be expressed
as a product of the matrix $A$ and the intrinsic flux $F^0(\nu_\alpha)$
\begin{displaymath}
\left[
\begin{array}{c}
F(\nu_e)\\
F(\nu_{\mu}) \\
F(\nu_{\tau})
\end{array}
\right]
\; = \;
A A^T \left[
\begin{array}{cc}
F^0(\nu_e)\\
F^0(\nu_{\mu})\\
F^0(\nu_{\tau})
\end{array}
\right] \; ,
\label{eq:flux}
\end{displaymath}
where $A$ is defined by (\ref{eq:transp})
\begin{displaymath}
\begin{array}{c}
A
\end{array}
\; \equiv \;
 \left[
\begin{array}{ccc}
|U_{e1}|^2 &|U_{e2}|^2 & |U_{e3}|^2 \\
|U_{\mu 1}|^2 &|U_{\mu 2}|^2 & |U_{\mu 3}|^2  \\
|U_{\tau 1}|^2 &|U_{\tau 2}|^2 & |U_{\tau 3}|^2
\end{array}
\right] \; .
\label{eq:AA}
\end{displaymath}
Since there is an experimental evidence \cite{CHOOZ} that $U_{e3} \simeq 0$, it means
that $\nu_{\mu}$ and $\nu_{\tau}$ are maximally mixed so that $|U_{\mu j}|^2-|U_{\tau j}|^2 \ll 1$ ($j=1,2,3$).
This symmetry between $\nu_{\mu}$ and $\nu_{\tau}$ results in the complete mixture of the all three neutrino species  which would arrive at Earth
in equal proportion \cite{AJY} $1:1:1$.

An interesting possibility of the neutrino decay has been considered recently \cite{BBHPW,TW} and  \cite{BQ}. The two body decays  have been considered
\be
\nu_i \rightarrow \nu_j + X {\;\;\rm and\;\;} \nu_i \rightarrow \bar{\nu_j} + X\; ,
\label{eq:nudecay}
\ee
where $X$ is some very light massless particle, for example a Majoron.
It has been shown \cite{BBHPW,TW,BQ} that the decays can lead to ratios which are dramatically different than $1:1:1$ resulting from oscillations. Thus an important experimental challenge is the flavour decomposition of the incoming neutrino flux. The deviation from the uniform mixture would definitely point towards new physics like unstable neutrinos or CPT violation.


\section*{Acknowledgments}
This work was
supported in part by Polish Committee for Scientific Research
 grant No. KBN 5P03B 14420. I thank Krzysztof Golec-Biernat for critically
reading the manuscript and useful comments.


\end{document}